\documentclass[twocolumn,letterpaper,prb,superscriptaddress,showpacs]{revtex4-1}
\usepackage{graphicx}
\usepackage{subfigure}
\usepackage{amsmath}
\usepackage{mathrsfs}
\DeclareRobustCommand{\orderof}{\ensuremath{\mathcal{O}}}

\begin{document}

\title{Anisotropic Multipolar Exchange Interactions in Systems with Strong
Spin-Orbit Coupling}
\author{Shu-Ting Pi}
\email{spi@ucdavis.edu}
\affiliation{Department of Physics, University of California, Davis, One Shields Avenus,
Davis, California 95616 USA}
\author{Ravindra Nanguneri}
\email{rnanguneri@ucdavis.edu}
\affiliation{Department of Physics, University of California, Davis, One Shields Avenus,
Davis, California 95616 USA}
\affiliation{Department of Chemistry and Biochemistry, University of Notre Dame, Notre
Dame, Indiana, 46556 USA}
\author{Sergey Savrasov}
\email{savrasov@physics.ucdavis.edu}
\affiliation{Department of Physics, University of California, Davis, One Shields Avenus,
Davis, California 95616 USA}
\date{\today}

\begin{abstract}
We introduce a theoretical framework for computaions of anisotropic
multipolar exchange interactions found in many spin--orbit coupled magnetic
systems and propose a method to extract these coupling constants using a
density functional total energy calculation. This method is developed using
a multipolar expansion of local density matrices for correlated orbitals
that are responsible for magnetic degrees of freedom. Within the mean--field
approximation, we show that each coupling constant can be recovered from a
series of total energy calculations via what we call the ``pair--flip''
technique. This technique flips the relative phase of a pair of multipoles
and computes corresponding total energy cost associated with the given
exchange constant. To test it, we apply our method to Uranium Dioxide, which
is a system known to have pseudospin $J=1$ superexchange induced dipolar,
and superexchange plus spin--lattice induced quadrupolar orderings. Our
calculation reveals that the superexchange and spin--lattice contributions
to the quadrupolar exchange interactions are about the same order with
ferro-- and antiferro--magnetic contributions, respectively. This highlights
a competition rather than a cooperation between them. Our method could be a
promising tool to explore magnetic properties of rare--earth compounds and
hidden--order materials.
\end{abstract}

\maketitle

\section{Introduction}

Solid--state systems with strong spin--orbit coupling have been a research
frontier for decades due to their rich magnetic phases that cannot be
explained by simplified model Hamiltonians. Among their peculiar properties,
the existence of multipolar moments may be one characteristics most
inaccessible to experimental investigation\cite{TR-1}. Interactions between
such moments not only induce complexity in high--rank magnetic order as
observed in $LaFeAsO$\cite{MO-1,MO-2}, $PrFe_{4}P_{12}$\cite{MO-3,MO-4}, $%
UPt_{3}$\cite{MO-5}, $YbRu_{2}Ge_{2}$\cite{MO-6}, $UO_{2}$\cite%
{PW-1,UE-1,UE-2,UE-3,UE-4} and many other compounds\cite{MO-7,MO-8} but also
exhibit the phenomena of hidden order phases as observed in $NpO_{2}$\cite%
{HO-1,HO-2,HO-9}, $Ce_{1-x}La_{x}B_{6}$\cite{HO-3} and $URu_{2}Si_{2}$\cite%
{HO-4,HO-5,HO-6,HO-7,HO-8}. Because of the active orbital degrees of
freedom, the conventional $S=1/2$ Heisenberg model\cite{SE-1} is no longer adequate to
describe their magnetic moments and, instead, high--rank tensor operators are required to form a complete basis\cite{TR-1}. The introduction of
multipolar moments makes the exchange interactions complicated, with a great
number of coupling constants, and makes their computation a difficult
problem in condensed matter physics.

Earlier studies of the exchange interactions in spin--orbital systems have
been developed by Coqblin and Schrieffer. They implemented the
Schrieffer--Wolff transformation on a spin--orbit coupled Anderson lattice
model and transformed it to a Kondo lattice problem so that the RKKY
interaction could be deduced\cite{CS-1,CS-2,CS-3,CS-4,CS-5}. Unlike
conventional $S=1/2$ Heisenberg model where the RKKY interaction is
isotropic\cite{MD-2}, the RKKY interaction for the spin--orbital model has an intrinsic
anisotropy even in a homogeneous system. In the 80's, Cooper et. al. solved
the Coqblin--Schrieffer Hamiltonian for $4f^{1}$ Cerium monopnictides and
explained their many unusual properties that conventional exchange models
failed to reproduce\cite{CP-1,CP-2,CP-3,CP-4,CP-5,CP-6}. In the 90's, they
further suggested a first--principles scheme to calculate the coupling
constants of a few simple materials and obtained satisfactory results\cite%
{CAB-1,CAB-2,CAB-3,CAB-4,CAB-5}. Although we now have a better understanding
about the multipolar exchange interactions nowadays, a comprehensive
physical picture remains lacking. Most of the models and computational
schemes are either based on the knowledge of specific exchange mechanisms or
too complicated to apply for materials. In this paper, we propose a method
to compute the multipolar coupling constants using a total energy electronic
structure calculation based on density functional theory (DFT) in its local
density approximation (LDA) or using an LDA+U approach\cite{MD-3}
. A short account of the present work has appeared already\cite{PW-1}.

We begin with a quick review of the RKKY interaction in spin--orbital
systems in Sec.II. These works were mostly contributed by Coqblin,
Schrieffer and Cooper and we emphasize the mechanism that induces the
intrinsic anisotropy of the exchange interactions. The formulation of
multipolar tensor operators is given in Sec.III. The language of multipolar
tensor operators is the most natural one to describe spin--orbit coupled
systems. Using this language, density matrices can be split into scalar,
dipolar, quadrupolar and higher multipolar components based on their
rotational symmetry. The complicated exchange coupling matrix may become
simplified and even diagonal when expressed in this tensor space. In Sec.IV,
an efficient method to deduce coupling constants using the LDA+U
electronic--structure calculation is introduced. We call this method the
``pair--flip technique'' because it relates every coupling parameter to a
series of total energy calculations by flipping the relative phase of a
multipole pair. Application to Uranium Dioxide ($UO_{2}$), is discussed in
Sec.V. $UO_{2}$ being famous for its important applications in nuclear
energy industry, is known to have pseudospin $J=1$ ground state, with both
dipolar and quadrupolar moment orderings, and therefore is a good candidate
to test our method. We find the superexchange contribution in $UO_{2}$ tends
to be ferromagnetic, which is very different from past studies. We conclude
in Sec.VI by speculating that our method could be a promising tool to
explore other spin--orbit coupled systems and materials with the hidden
order.  
\begin{figure}[tbp]
\centering
\includegraphics[width=0.8\columnwidth]{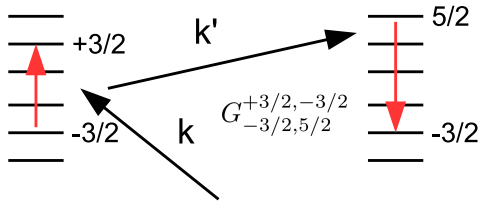}
\caption{(color) The RKKY mechanism of a $J=5/2$ system: an incoming free
electron with crystal momentum $k$ interacts with a local moment and induces
a transition between the degenerate states $-3/2$ to $3/2$. Then it leaves
with $k^{^{\prime }}$ and induces another transition $5/2$ to $-3/2$ at a
neighboring site. These transitions are coupled by the exchange constant $%
G_{-3/2,5/2}^{+3/2,-3/2}$}
\end{figure}

\section{Spin-Orbit Coupled Exchange Interactions}

\subsection{Exchange Interactions}

Exchange interaction appears in an effective model for an Anderson lattice
Hamiltonian in its low excitation limit where the particle fluctuation is
frozen and only transitions among the internal degrees of freedom, $i.e.$ the
degenerate single particle states, are allowed. The Anderson lattice model%
\cite{MD-1,MD-4} is given by 
\begin{align}
H& =\sum_{\mathbf{k}\sigma }\epsilon _{\mathbf{k}\sigma
}+\sum_{d}\{\epsilon _{d}n_{d \sigma }+Un_{d \uparrow }n_{d \downarrow }\} \\
& +\sum_{\mathbf{k} d \sigma }\{V_{\mathbf{k} d}c_{\mathbf{k} \sigma
}^{\dagger }c_{d \sigma }+h.c\},  \notag
\end{align}%
where $d$ is the localized correlated state, $\epsilon _{d}$ is the on--site
energy of the localized $d$ orbital, $\mathbf{k}$ is the crystal momentum, $%
\sigma $ is the spin index, $U$ is the Hubbard interaction, $V$ is the
coupling between a conduction electron and a localized $d$ state. Let us
denote the first two terms as $H_{0}$ and the last one as $H_{1}$. In the
Kondo limit $U\gg \epsilon _{d}$, charge transfer is frozen and the Anderson
lattice model becomes the Kondo lattice model. In the 60's, Schrieffer and
Wolff suggested a procedure to eliminate the charge fluctuation effects\cite%
{CS-1} (high order perturbation of $H_{1}$) by introducing a unitary
transformation that keeps $H$ to $\orderof(H_{1})$ only, $H^{^{\prime
}}=e^{s}He^{-s}\sim \orderof(H_{1})$. It requires $[H_{0},s]=H_{1}$ and $%
H^{^{\prime }}=H_{0}+\frac{1}{2}[s,H_{1}]$. Coqblin and Schrieffer
implemented this transformation to a spin--orbit coupled $J=5/2$ Cerium ($%
4f^{1}$) system and derived the spin--orbital version of RKKY interaction
which describes the exchange interaction between the two local moments\cite%
{CS-2,CS-3}.

A general form of the two--ion exchange interaction in a spin--orbit coupled
lattice system can be written as\cite{TR-1} 
\begin{equation*}
H(J)=\sum_{i j}\sum_{\alpha  \beta  \gamma  \delta }G_{\gamma  \alpha
}^{\delta  \beta }(i, j)c_{i \delta }^{\dagger }c_{i \gamma }c_{j \beta
}^{\dagger }c_{j \alpha },
\end{equation*}%
where $i,j$ are site indices, $\alpha ,\beta ,\gamma ,\delta $ are labels of
the degenerate states which range from $-J\ $to $J$, $J$ being the quantum
number of the total moment. The physics of this Hamiltonian is easy to
understand (see Fig.1). It describes the transition from $\alpha $ to $\beta 
$ at site $i$ and another transition from $\gamma $ to $\delta $ at site $j$%
. These transitions are coupled by a constant $G_{\gamma \alpha }^{\delta
\beta }(i,j)$. There are many mechanisms to induce these transitions, $e.g.$
RKKY (interaction with conduction electrons), superexchange (interaction
with neighboring non--magnetic atoms), and spin--lattice coupling
(interaction with lattice vibrations)\cite{MD-2}. 
\begin{figure}[tbp]
\centering
\includegraphics[width=0.8\columnwidth]{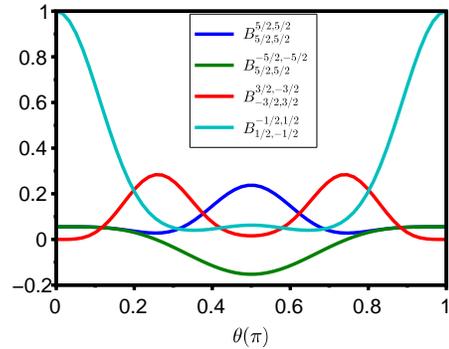}
\caption{(color) The angular dependence of the coupling constants $B_{%
\protect\gamma \protect\alpha }^{\protect\delta \protect\beta }(\protect%
\theta, \protect\phi )$ as a function of $\theta$ (in unit of $\pi$) with $\protect\phi =0$. Anisotropy can be found in
all transitions.}
\end{figure}

\subsection{Anisotropy}

A major feature of the spin--orbit coupled exchange interaction is its
anisotropy. To show this, let us consider a simple but still realistic model
($e.g.$ Cerium compounds) that each site has $f^{1}$ configuration with $J=5/2$
ground state. As given by Ref.\cite{CP-2}, the coupling matrix induced by
the RKKY mechanism has the following form: 
\begin{equation*}
H=\sum_{ij}E(|r_{ij}|)\sum_{\alpha \beta \gamma \delta }B_{\gamma
\alpha }^{\delta \beta }(\theta, \phi )L_{\delta \gamma }^{i}L_{\beta
\alpha }^{i},
\end{equation*}%
with 
\begin{align}
B_{\gamma \alpha }^{\delta \beta }(\theta, \phi )& =e^{i(\delta -\gamma
+\beta -\alpha )\phi }\times  \\
\sum_{MM^{\prime }=\pm 1/2}& \{d_{\delta M^{^{\prime }}}(\theta )d_{\gamma
M}(\theta )-\frac{1}{6}\delta _{MM^{^{\prime }}}\delta _{\delta \gamma }\} 
\notag \\
\times & \{d_{\beta M}(\theta )d_{\alpha M^{^{\prime }}}(\theta )-\frac{1}{6}%
\delta _{M M^{^{\prime }}}\delta _{\beta \alpha }\},  \notag
\end{align}%
where $L_{\delta \gamma }^{i}=|\delta ^{i}\rangle \langle \gamma ^{i}|$ is
the transition operator which is the single particle version of $c_{i\delta
}^{\dagger }c_{i\gamma }$, $d_{\beta M}(\theta )$ is the quantum mechanical
rotation matrix of $J=5/2$. Some matrix elements of the function $B_{\gamma
\alpha }^{\delta \beta }(\theta, \phi )$ as a function of angle are shown
in Fig.2.

\begin{figure}[tbp]
\centering
\includegraphics[width=1.0\columnwidth]{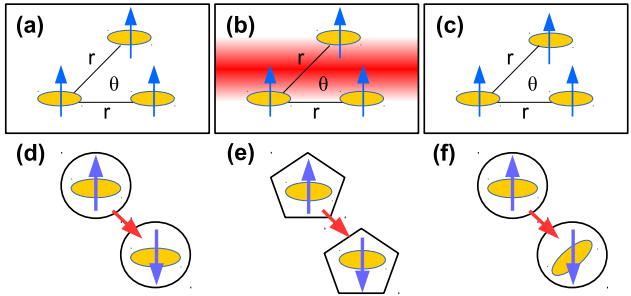}
\caption{(color) Anisotropy of exchange interactions. The blue arrows are
local moments and the yellow ellipses are orbitals. (a) S=1/2 in a
homogeneous background (b) S=1/2 in an inhomogeneous background. The
anisotropy can be induced by many factors, $e.g.$ crystal structure,
electronic structure and external fields. (c) J=5/2 in a homogeneous
background. In (d)--(f), we use a cage around the moment to represent the
relation between the background and the local moment as described by
(a)--(c) respectively: (d) Transition of the spin moment in a homogeneous
background will not change anything; the exchange interaction is isotropic,
(e) Transition of the moment in an inhomogeneous background gives a
different configuration; the exchange interaction is anisotropic (f)
Transition of a spin--orbit coupled moment in a homogeneous background:
Although the background has no directional dependence, due to highly
anisotropic shapes of the orbitals exchange interaction becomes
directionally dependent.}
\end{figure}

Unlike conventional $S=1/2$ RKKY where the coupling matrix has only $%
E(|r_{ij}|)$ dependence\cite{MD-2}, one can immediately find that all the matrix
elements are highly anisotropic for the spin--orbit coupled systems. For
example, transitions $[\frac{1}{2}\rightarrow -\frac{1}{2}]$ and $[-\frac{1}{%
2}\rightarrow \frac{1}{2}]$ are strongly coupled only when two ions have
relative angle $\theta =0$ or $\pi $ and become almost decoupled when $%
\theta =\pi /4\sim \pi /3$. The physical origin of the intrinsic anisotropy
comes from the spatial dependence of atomic orbitals. In Fig.3, we consider
an exchange problem in a homogeneous and in an inhomogeneous systems. In (a)
and (d), since the background (for RKKY, the background is the sea of
conduction electrons) and the transition (varying the spin) are both
homogeneous, so the exchange interaction is isotropic. In (b) and (e), the
background is inhomogeneous, so a homogeneous transition still feels its
environmental anisotropy. In (c) and (f), even though the system is
homogeneous, the transition ($e.g.$ $-3/2$ to $1/2$) itself is always
anisotropic due to its coupling with the spatial wave function (recall that $%
f$--orbitals have highly anisotropic shapes). The anisotropy induced by the
active orbital degrees of freedom distinguishes the nature between a
spin--only and a spin--orbit coupled exchange interaction and makes the
calculation of the exchange matrix difficult due to the presence of many
off--diagonal matrix elements.

\section{Multipolar Tensor Operators}

In the following, we use a single--particle description while extension to a
many--body version can be achieved by replacing the ket and bra vectors by
creation and annihilation operators.

\subsection{Super Basis}

An unit transition tensor operator in the total moment $J$ Hilbert space is
defined as: 
\begin{equation*}
L_{\delta \gamma }(J)=|\delta \rangle \langle \gamma |,
\end{equation*}%
where $\delta ,\gamma $ are the magnetic moment states which range from $-J\ 
$to $+J$. A matrix defined in the same Hilbert can be expressed in terms of
the above operators, $e.g.$, for $J=1/2$ we deal with 2x2 matrices and their
expansions in terms of $L_{\delta \gamma }$: 
\begin{equation*}
\left( 
\begin{array}{cc}
1 & 2 \\ 
3 & 4%
\end{array}%
\right) =1\left( 
\begin{array}{cc}
1 & 0 \\ 
0 & 0%
\end{array}%
\right) +2\left( 
\begin{array}{cc}
0 & 1 \\ 
0 & 0%
\end{array}%
\right) +3\left( 
\begin{array}{cc}
0 & 0 \\ 
1 & 0%
\end{array}%
\right) +4\left( 
\begin{array}{cc}
0 & 0 \\ 
0 & 1%
\end{array}%
\right) .
\end{equation*}%
Using the language above, we have: 
\begin{equation*}
A=1L_{11}+2L_{12}+3L_{21}+4L_{22},
\end{equation*}%
where the coefficients can be obtained by taking the trace of the matrix and
the daggered tensor operator : $Tr[AL_{ij}^{\dagger }]$. This shows that $%
\{L_{\delta \gamma }(J)\}$ forms a basis with trace inner product and any
operator defined in the same Hilbert space can be expanded in terms of this
basis. In the following, we name the basis set that expands an operator
defined in the $J$ Hilbert space with trace inner product as
\textquotedblleft super basis\textquotedblright\ to distinguish from the
commonly used vector basis $\{|\delta \rangle \}$. The transition super
basis is not only the option. A spherical harmonics super basis can be
generated by\cite{TR-1} 
\begin{align}
Y_{K}^{Q}(J)& =\sum_{MM^{\prime }}(-1)^{J-M}(2K+1)^{2} \\
& \times \left( 
\begin{array}{ccc}
J & J & K^{^{\prime }} \\ 
M^{^{\prime }} & M & Q%
\end{array}%
\right) |JM\rangle \langle JM^{^{\prime }}|,  \notag
\end{align}%
where the parentheses denote a $3j$--symbol; $K$ is the rank which ranges $%
0\sim 2J$; $Q$ is the projection index of rank $K$ which ranges from $-J\ $%
to $+J$. Similarly a matrix defined in the $J$ Hilbert space has the
property $A(J)=\sum_{K Q}\alpha _{K}^{Q}Y_{K}^{Q}(J)$ and the expansion
coefficients can be calculated $\alpha _{K}^{Q}=Tr[AY_{K}^{\dagger Q}]$. One
can easily verify that there are $(2J+1)^{2}$ members in the spherical
harmonics super basis which is exactly the number of matrix elements (also
the number of members in the transition super basis) in the $J$ Hilbert
space.

For the same example of $J=1/2$ we have 
\begin{align*}
\left( 
\begin{array}{cc}
1 & 2 \\ 
3 & 4%
\end{array}%
\right) & =\frac{5}{2}\left( 
\begin{array}{cc}
1 & 0 \\ 
0 & 1%
\end{array}%
\right) +2\left( 
\begin{array}{cc}
0 & 1 \\ 
0 & 0%
\end{array}%
\right)  \\
& +\frac{3}{2}\left( 
\begin{array}{cc}
-1 & 0 \\ 
0 & 1%
\end{array}%
\right) -3\left( 
\begin{array}{cc}
0 & 0 \\ 
-1 & 0%
\end{array}%
\right) .
\end{align*}%
Here, the spherical harmonics super basis is actually the unit, the $z$%
--projection, and the ladder (raising and lowering) operators with
appropriate normalization constants: $Y_{0}^{0}\sim I$, $Y_{1}^{0}\sim
\sigma ^{z}$, $Y_{1}^{+1}\sim \sigma ^{x}+i\sigma ^{y}$, $Y_{1}^{-1}\sim
\sigma ^{x}-i\sigma ^{y}$. It is called the spherical harmonics super basis
because its members follow exactly the same symmetry as the spherical
harmonics. $Y_{0}^{0}$ behaves like a $s$--orbital; $Y_{1}^{-1}$, $Y_{1}^{0}$
and $Y_{0}^{+1}$ behave like $p^{-1}$, $p^{0}$ and $p^{+1}$ orbitals.

In group theory, these operators are named after their rank: $K=0$ scalar, $%
K=1$ dipole, $K=2$ quadrupole, $K=3$ octupole, etc. We have to emphasize
that the multipoles in this framework are different form those in the theory
of electromagnetism, where the multipoles refer to the spatial distribution
of charge $\rho (\mathbf{r})$ or magnetization $\mathbf{m}(\mathbf{r})$
expanded by multipolar functions $Y_{l}^{m}(\theta ,\phi )$. Here, the
multipoles do not refer to any spatial distribution but to the rotational
properties of a matrix, or more precisely, to the transitions of magnetic
moments. Although they follow the same algebra, they do not correspond to
the same physical meaning.

Similarly, we can also define the cubic harmonics super basis, where all the
operators are Hermitian\cite{TR-1} 
\begin{align}
T_{K}^{Q}& =\frac{1}{\sqrt{2}}[(-1)^{Q}Y_{K}^{Q}(J)+Y_{K}^{-Q}(J)] \\
T_{K}^{-Q}& =\frac{i}{\sqrt{2}}[Y_{K}^{-Q}(J)-(-1)^{Q}Y_{K}^{-Q}(J)].  \notag
\end{align}%
For $J=1/2$ case, these are Pauli matrices: $T_{0}^{0}\sim I$, $%
T_{1}^{+1}\sim \sigma _{y}$, $T_{1}^{0}\sim \sigma _{z}$ and $T_{0}^{0}\sim
\sigma _{x}$. Also, this basis follows the same symmetry as cubic harmonics: 
$s$, $p^{x}$, $p^{y}$ and $p^{z}$. Therefore, instead of using abstract $%
(K,Q)$ indexes, we will label these tensor operators using their symmetry: $%
T^{s}$, $T^{x}$, $T^{y}$, $T^{z}$, $T^{xy}$, $T^{yz}$, $T^{zx}$, etc. It is
straightforward to rewrite the spin--orbit coupled exchange interaction by
using different super bases: 
\begin{align*}
H& =\sum_{ij}\sum_{\alpha \beta \gamma \delta }G_{\gamma \alpha
}^{\delta \beta }(i,j)L_{\delta \gamma }^{i}(J)L_{\beta \alpha }^{i}(J) \\
& =\sum_{ij}\sum_{KQ}F_{K_{i}K_{j}}^{Q{i}%
Q_{j}}Y_{Q_{i}}^{K_{i}}(J)Y_{Q_{j}}^{K_{j}}(J) \\
& =\sum_{ij}\sum_{KQ}C_{K_{i}K_{j}}^{Q{i}%
Q_{j}}T_{Q_{i}}^{K_{i}}(J)T_{Q_{j}}^{K_{j}}(J).
\end{align*}%
The couplings among multipolar operators now appear naturally, and the
coupling matrices in different bases can be linked by using unitary
transformations. Obviously, one can define other super bases by making
different linear combinations of them. The benefit of using appropriate
super basis is that the coupling matrix may become block diagonal or even
completely diagonal when using appropriate symmetry. 
\begin{figure}[tbp]
\centering
\includegraphics[width=1.0\columnwidth]{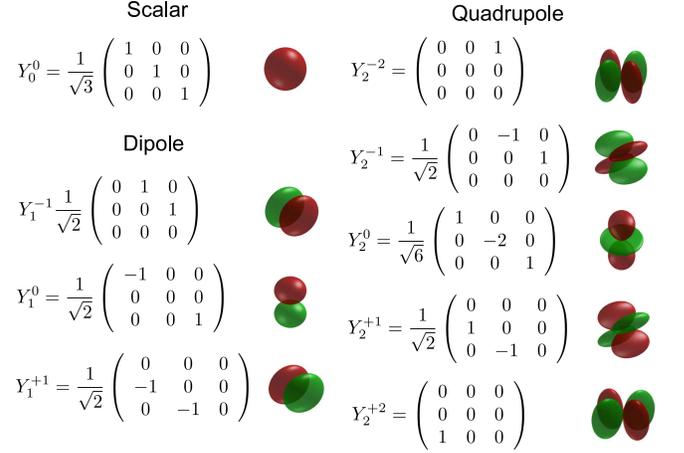}
\caption{(color) The spherical tensor operators of $J=1$ and the real part
of analogous spherical harmonics. States to expand the matrices are ordered
by $|-1\rangle $, $|0\rangle $ and $|+1\rangle $. Each member has its
analogous spatial function as $s$, $p^{-1}$, $p^{0}$, $p^{+1}$, $d^{-2}$, $%
d^{-1}$, $d^{0}$, $d^{+1}$ and $d^{+2}$ spherical harmonics which follow the
same symmetry under rotation.}
\end{figure}

\subsection{Physical Pictures}

To illustrate the physics of multipolar tensor operators, let us focus on
the spherical harmonics operators for $J=1$. In this case, we have 9
linearly independent tensor operators as shown in Fig.4. We also display the
shapes of $s$, $p$ and $d$ spherical harmonics functions to represent their
analogy. Because the rotational properties of those tensor operators are the
same as the original spherical harmonics, we can ``
visualize'' these matrices by this way.

Let us discuss the scalar term first. The scalar term is exactly an identity
matrix, and since the identity is invariant under rotations, it always looks
the same under any rotation as the $s$--orbital. An important feature of the
scalar term is its relation to the total charge. If we expand the density
matrix $\rho =\sum_{K,Q}\alpha _{K}^{Q}Y_{K}^{Q}$, the total charge of the
system is proportional to the coefficient $\alpha _{0}^{0}$. As for the
dipole terms, the matrices are no longer unchanged under rotation: $%
Y_{1}^{-1}$ and $Y_{1}^{+1}$ describe time--reversed transition processes
which change a single quantum of the angular momentum. $Y_{1}^{0}$ is
another diagonal matrix which induces no change of moment. Similar
descriptions also hold for quadrupoles: $Y_{2}^{-2}$ and $Y_{2}^{+2}$ change
two moment quanta; $Y_{2}^{-1}$ and $Y_{2}^{+1}$ change one moment quantum; $%
Y_{2}^{0}$ changes no moment. Although $Y_{1}^{\pm 1}$ and $Y_{2}^{\pm 1}$
both change one moment quantum, they are essentially different. Notice that
the non--zero matrix elements of $Y_{2}^{\pm 1}$ have a sign difference but $%
Y_{1}^{\pm 1}$ have no such term. If single quantum transition channels are
in--phase, it is a dipole; if they are out--of--phase, it is a quadrupole.
Similar analysis can be applied to other super bases. A diagrammatic
interpretation of the dipole--dipole and quadrupole--quadrupole RKKY
exchange interactions is shown in Fig.5.

\begin{figure}[tbp]
\centering
\includegraphics[width=1.0\columnwidth]{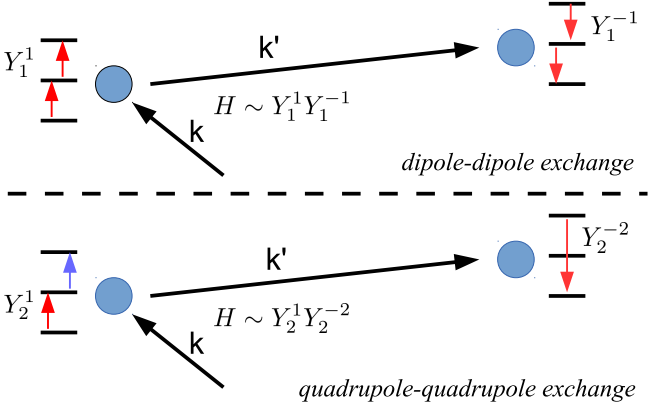}
\caption{(color) Dipole--dipole and quadrupole--quadrupole exchange
interactions. An incoming conduction electron with momentrum $k$ interacts
with a $J=1$ local moment, induces a $Y_{1}^{1}$ ($Y_{2}^{1}$) transition
and leaves with momentum $k^{^{\prime }}$ inducing a $Y_{1}^{-1}$ ($%
Y_{2}^{-2}$) transition on a neighboring site. The blue transition arrow for 
$Y_{2}^{1}$ means a negative phase compared to the red one.}
\end{figure}

\section{Method of Computing Coupling Matrix}

Here we introduce an efficient method to calculate the coupling matrix of a
multipolar exchange interaction using total energy electronic structure
calculation, such as LDA or LDA+U \cite{MD-3}. The discussion is
based on a specific total moment $J$, therefore labeling by $J$ will be
omitted.

\subsection{Energy Variation}

Consider a multipolar exchange interaction within the mean--field
approximation, and the ground state energy $E_{0}$: 
\begin{align}
H& =\sum_{ij}\sum_{KQ}C_{K_{i}K_{j}}^{Q{i}%
Q_{j}}T_{Q_{i}}^{K_{i}}T_{Q_{j}}^{K_{j}}  \notag \\
& \simeq \sum_{ij}\sum_{KQ}C_{K_{i}K_{j}}^{Q{i}Q_{j}}\{\langle
T_{Q_{i}}^{K_{i}}\rangle T_{Q_{j}}^{K_{j}}+T_{Q_{i}}^{K_{i}}\langle
T_{Q_{j}}^{K_{j}}\rangle \},  \notag \\
E_{0}& =\langle H\rangle =2\sum_{ij}\sum_{KQ}C_{K_{i}K_{j}}^{Q{i}%
Q_{j}}\langle T_{Q_{i}}^{K_{i}}\rangle \langle T_{Q_{j}}^{K_{j}}\rangle .
\end{align}%
The formula for the ground state energy is exactly the classical version of
the multipolar exchange interaction. In this case, the multipolar moments
are no longer quantized and can vary continuously. If we introduce
variations for a particular pair of multipoles at different sites: 
\begin{align*}
\langle T_{Q_{i}}^{K_{i}}\rangle & \rightarrow \langle
T_{Q_{i}}^{K_{i}}\rangle +\delta \langle T_{Q_{i}}^{K_{i}}\rangle , \\
\langle T_{Q_{j}}^{K_{j}}\rangle & \rightarrow \langle
T_{Q_{j}}^{K_{j}}\rangle +\delta \langle T_{Q_{j}}^{K_{j}}\rangle ,
\end{align*}%
plug them into the formula for $E_{0}$ , we obtain: 
\begin{equation*}
C_{K_{i}K_{j}}^{Q{i}Q_{j}}(i,j)=\frac{1}{2}\frac{\delta
^{2}E_{K_{i}K_{j}}^{Q{i}Q_{j}}}{\delta \langle T_{K_{i}}^{Q_{i}}\rangle
\delta \langle T_{K_{j}}^{Q_{j}}\rangle },
\end{equation*}%
and 
\begin{align*}
\delta ^{2}E_{K_{i}K_{j}}^{Q{i}Q_{j}}& =(\delta E_{K_{i}K_{j}}^{Q{i}%
Q_{j}}-\delta E_{K_{i}}^{Q{i}}-\delta E_{K_{j}}^{Q{j}}), \\
\delta E_{K_{i}}^{Q{i}}& =E_{K_{i}}^{Q{i}}-E_{0},
\end{align*}%
where $E_{K_{i}}^{Q{i}}$ is the new energy associated with a variation $%
\delta \langle T_{Q_{i}}^{K_{i}}\rangle $ of $\langle
T_{Q_{i}}^{K_{i}}\rangle $ multipole. Therefore, in order to calculate a
multipolar exchange constant we need to obtain three energies: the energy
cost of making a variation on site $i$, the energy cost of making a
variation on site $j$ and the energy cost of making the same variations on
both sites. If the multipoles $T_{Q_{i}}^{K_{i}}$ and $T_{Q_{j}}^{K_{j}}$
are not coupled, the energy cost of varying both will be simply the sum of
two independent variations on each site. However, if they are coupled,
varying both sites simultaneously will induce an extra exchange energy which
is proportional to the exchange constant as shown in Fig.6. Therefore, if
one can compute these energies using advanced electronic structure
calculation, one is able to obtain the effective exchange interaction.

Another issue is how to perform a variation on the multipoles in a realistic
calculation. To answer this question, we have to use the trace inner product
theorem. Consider the local density matrices of each correlated site. The
local density matrices can be expanded by a super basis defined on that
site: $\rho _{i}=\sum_{K_{i},Q_{i}}\alpha _{K_{i}}^{Q_{i}}Y_{K_{i}}^{Q_{i}}$%
. We intentionally choose a super basis where all the tensor operators are
Hermitian, $e.g.$, cubic harmonics, so we also have: $\langle
T_{K_{i}}^{Q_{i}}\rangle =Tr[\rho T_{K_{i}}^{Q_{i}}]=Tr[\rho
T_{K_{i}}^{\dagger Q_{i}}]=\alpha _{K_{i}}^{Q_{i}}$. It means $\langle
T_{K_{i}}^{Q_{i}}\rangle \rightarrow \langle T_{K_{i}}^{Q_{i}}\rangle
+\delta \langle T_{K_{i}}^{Q_{i}}\rangle $ is essentially $\alpha
_{K_{i}}^{Q_{i}}\rightarrow \alpha _{K_{i}}^{Q_{i}}+\delta \alpha
_{K_{i}}^{Q_{i}}$. Therefore, we can vary a multipole by changing its
corresponding expansion coefficient. 
\begin{figure}[tbp]
\centering
\includegraphics[width=1.0\columnwidth]{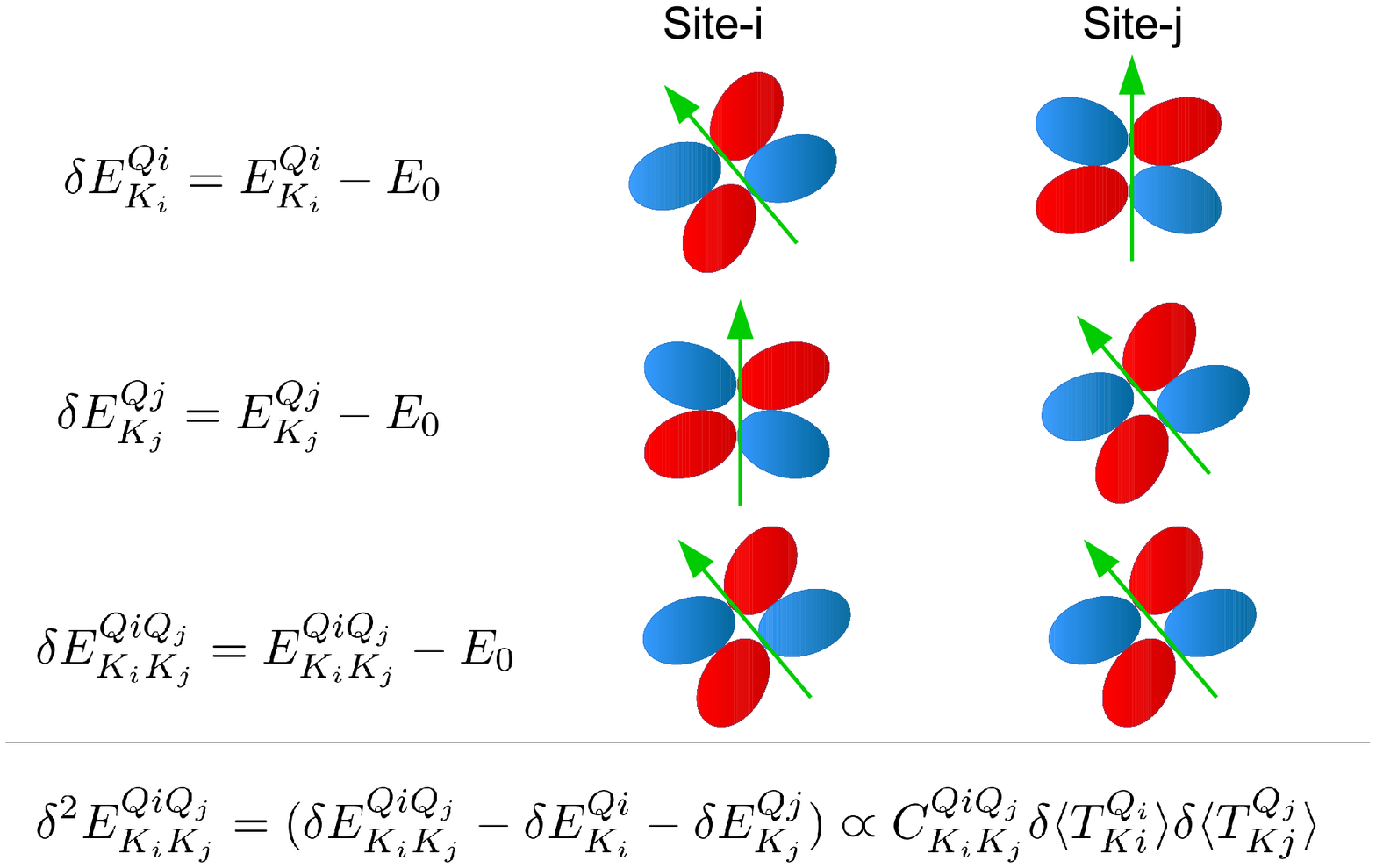}
\caption{(color) The meaning of each term in computing coupling constants. 
Here we use $T^{yz}$ to represent the multipoles.}
\end{figure}

\subsection{Pair--flip Technique}

When mapping the exchange model Hamiltonian to a series of total energy
calculations using our method, it is necessary to make sure the total
energies only contain the contributions from exchange interactions. This is
not very straightforward. Recall that, in $S=1/2$ Heisenberg model, we
relate the exchange constants to the new total energy of the meta--stable
state with one or more spin moments flipped. Therefore when calculating the
total energy, one should not perform any self--consistent calculation of the
tensor flipped configuration else the system may evolve and go back to its
ground state. To avoid the latter, we use the Anderson force theorem\cite%
{AF-1} and read the band energy differences only, $i.e.$ the energies
associated with occupied single--particle states. We also apply several
constraints on the variations: 1) keep the total charge conserved; 2) keep
the symmetry to avoid crystal field effects; 3) keep the magnitude of the
multipolar moments fixed; 4) enforce the hermitianness \ of the density
matrix. Combining these constraints, the only possible choice is to add a
phase on the expansion coefficients $\alpha _{K_{i}}^{Q_{i}}$ of the density
matrix $\rho _{i}=\sum_{K_{i}Q_{i}}\alpha _{K_{i}}^{Q_{i}}Y_{K_{i}}^{Q_{i}}$
and the simplest one is a minus sign: $\alpha _{K_{i}}^{Q_{i}}\rightarrow
-\alpha _{K_{i}}^{Q_{i}}$. When this is done, $\delta \alpha
_{K_{i}}^{Q_{i}}=-2\alpha _{K_{i}}^{Q_{i}}$. This is similar to the way we
calculate the exchange constants in conventional $S=1/2$ Heisenberg model, 
\textit{i.e.} relating the exchange constants to the energy cost of flipping
a spin moment (changing the sign of the $z$--axis spin projection).

However, the term \textquotedblleft flip\textquotedblright\ has a different
meaning in the language of multipolar exchange interactions from the case of 
$S=1/2$. In the conventional $S=1/2$ Heisenberg model, a ``
flip'' means the flipping of the local axis of a spin moment. For most
multipoles, such a flip is meaningless because it generates no change.
Instead, the most general concept of a `` flip'' is to put a
minus sign on their expansion coefficients or, equivalently, flip their
phase ( a $\pi $ phase gives us $e^{i\pi }=-1$). In Fig.7, we show the
pictures of a \textquotedblleft phase flip" for cubic harmonics tensor
operators via changing the sign on their corresponding functions. For
dipoles, it is indeed equivalent to flipping its local axes. However, for
quadrupoles, a phase flip of $T^{xy}$, $T^{yz}$, $T^{zx}$ and $%
T^{x^{2}-y^{2}}$ is actually a $\pi /2$ rotation along different axes, and
for $T^{3z^{2}}$, it cannot be characterized as a rotation. Therefore, when
we say \textquotedblleft antiferromagnetic\textquotedblright\ $T^{yz}$
quadrupolar ordering, it actually means the $T^{yz}$ quadrupoles are ordered
by a $\pi /2$ rotation alternatively\cite{UE-1} or, more precisely, by a $%
\pi $ phase change.

Now, we summarize how to calculate the exchange constants using the
pair--flip technique: 1) Calculate the ground state of the system using
advanced electronic structure calculation, such as LDA or LDA+U. 2) Generate
an appropriate super basis that is consistent with the symmetry of the
system. 3) Expand the local density matrices of the magnetic orbitals in
this super basis. 4) Pick a pair of tensor components on different sites,
flip their phases separately and simultaneously, recombine them into new
local density matrices (make sure they are still Hermitian) and calculate
their corresponding band energies (making sure not to do any
self--consistent calculation on these meta--stable states). 6) Read the band
energies and calculate the exchange coefficients. 
\begin{figure}[tbp]
\centering
\includegraphics[width=1.0\columnwidth]{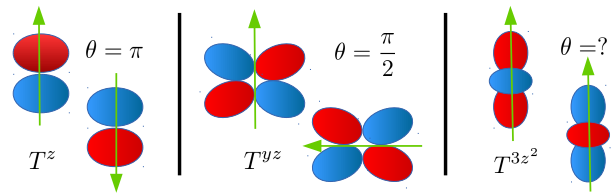}
\caption{(color) Phase flip of multipoles. A phase flip is different form
the axis flip. For $T^{yz}$, it is a $\protect\pi /2$ rotation along the $x$
axis. For $T^{3z^{2}}$, it does not correspond to any rotation. Only dipoles
fit the concept of flipping their axes.}
\end{figure}

\section{Application to Uranium Dioxide}

\subsection{Model Hamiltonian}

To test our method, we use Uranium Dioxide (UO$_{2}$) as a benchmark
material due to the presence of dipolar and quadrupolar order in its ground
state. UO$_{2}$ has been one of the widely discussed actinide compounds due
to its applications in nuclear energy industry. It is a Mott insulator with
cubic structure and well--localized $5f^{2}$ electrons (Uranium valence is U$%
^{4+}$ by naive charge counting). Below $T_{N}=30.8K$ it undergoes a
first--order magnetic and structural phase transition where a non--collinear
antiferromagnetic (AFM) phase with transverse 3--$\mathbf{k}$ magnetic
ordering accompanied by the cooperative Jahn--Teller distortion occurs \cite%
{UE-2}. The two--electron ground state forms a $\Gamma _{5}$ triplet,
holding pseudospin $J=1$ rotation symmetry\cite{UE-1} 
\begin{align*}
|+\rangle & =\sqrt{\frac{7}{8}}|+3\rangle -\sqrt{\frac{1}{8}}|-1\rangle
;\quad \langle +|J_{z}|+\rangle =\frac{5}{2}, \\
|0\rangle & =\sqrt{\frac{1}{2}}|+2\rangle -\sqrt{\frac{1}{2}}|-2\rangle
;\quad \langle 0|J_{z}|0\rangle =0, \\
|-\rangle & =-\sqrt{\frac{7}{8}}|-3\rangle +\sqrt{\frac{1}{8}}|+1\rangle
;\quad \langle -|J_{z}|-\rangle =-\frac{5}{2}.
\end{align*}%
The numbers in the kets of the right--hand--side label the $m_{J}$ of the $%
^{3}H_{4}$ configuration. It makes $UO_{2}$ a good choice to test our
method, as it is a minimal challenge beyond $S=1/2$ Heisenberg model.

As discussed in the previous sections, the description of a spin--orbit
coupled $J=1$ system requires the existence of dipolar and quadrupolar
moments. It is commonly believed that there are two major mechanisms to
induce exchange coupling in this system: 1) superexchange (SE), and 2)
spin--lattice interaction (SL). The former contributes to both dipole and
quadrupole and the latter contributes to quadrupole only due to the symmetry
of structural lattice distortion. The dominance of SE or SL in affecting the
quadrupole exchange remains a controversial issue \cite{UE-1,UE-2,UE-3,UE-4}%
. Since our method is based on a static electronic structure calculation, we
do not explore dynamical effects in all their details. Therefore, separate
calculations using the coupled frozen--phonon and frozen--magnon techniques
were performed to extract the SL coupling constants.

\begin{figure}[tbp]
\centering
\includegraphics[width=0.8\columnwidth]{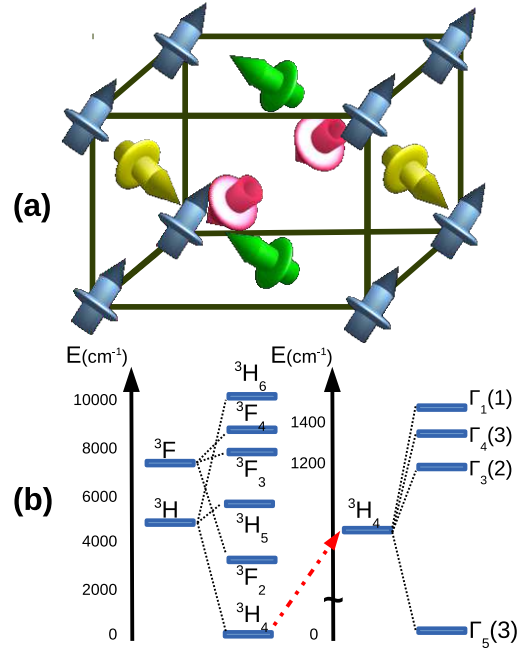}
\caption{(color) (a) Magnetic moments of dipoles (arrows) and quadrupoles
(disk) in the 3-$\mathbf{k}$ structure. (b) The energy splitting of low
lying states in $UO_{2}$ \protect\cite{UE-2}. The $^{3}H$ and $^{3}F$ states
of free $U^{4+}$ ion are split into $^{3}H_{4}$ multiplets and other excited
states by spin--orbit coupling and further split into the $\Gamma _{5}$
triplet ground state by crystal fields. Their degeneracy are shown inside the
parentheses}
\end{figure}

Since $UO_{2}$ is a cubic system, it is natural to take cubic harmonics as
our super basis: $T^{s}$ for rank 0; $T^{x}$, $T^{y}$, $T^{z}$ for rank $1$
(dipole); $T^{xy}$, $T^{yz}$, $T^{zx}$, $T^{x^{2}-y^{2}}$, $T^{3z^{2}}$ for
rank $2$ (quadrupole) \cite{TR-1,UE-1,UE-3}. The ground state local density
matrix of an U ion can be expanded by them $\rho _{i}=\sum_{m}\alpha
_{i}^{m}T_{i}^{m}$, where $i$ is site index, $m$ is the projection index for
cubic harmonics, and $\alpha _{i}^{m}=Tr[\rho _{i}T_{i}^{\dagger m}]$ is the
expansion coefficient. Since the triplet degeneracy of $\Gamma _{5}$ is
further split below $T_{N}$, we can approximate the ground state as $%
|GS\rangle =|-1\rangle $, the lowest energy state of an isolated U--ion in
the 3--$\mathbf{k}$ magnetic phase. 3--$\mathbf{k}$ ordering requires the
four $U$ sublattice moments to point in inequivalent $(1,1,1)$ directions,
which means the $|-1\rangle $ states are defined in different local
coordinates for each $U$ sublattice \cite{UE-2}. Thus, we need to make a
rotation on each site to ensure everything is in a global coordinate system.

In the local coordinate system, the expansion of density matrices has the
same tensor expansion coefficients: 
\begin{equation*}
\rho _{i}=|-1\rangle \langle -1|=\sqrt{\frac{1}{3}}T_{i}^{s}-\sqrt{\frac{1}{2%
}}T_{i}^{z}+\sqrt{\frac{1}{6}}T_{i}^{3z^{2}}
\end{equation*}%
When converting to global coordinate system, one has to apply a rotation
matrix $D(\theta _{i},\phi _{i},\psi _{i})^{\dagger }\rho _{i}D(\theta
_{i},\phi _{i},\psi _{i})$ using different Euler angles $(\theta _{i},\phi
_{i},\psi _{i})$ for each site. Then non--vanishing components of the ground
state 3--$\mathbf{k}$ quadrupolar order are $s$, $x$, $y$, $z$, $xy$, $yz$
and $zx$. Thus the model Hamiltonian of nearest--neighbor exchange
interaction between magnetic $U$ atoms is assumed to be (in the global
system): 
\begin{align}
h^{EX}& =h^{SE}+h^{SL} \\
&
=\sum_{m i j}C_{ij}^{mm}T_{i}^{m}T_{j}^{m}+%
\sum_{n i j}K_{ij}^{nn}T_{i}^{n}T_{j}^{n}  \notag \\
& m\in x,y,z,xy,yz,zx\quad ;\quad n\in xy,yz,zx,  \notag
\end{align}%
where ($i,j$) are the nearest--neighbor site indexes and ($C_{ij}^{mm}$, $%
K_{ij}^{nn}$) are the exchange constants from SE and SL respectively.
Couplings between tensor operators with different symmetry indexes are
prohibited by cubic symmetry. This fact demonstrates the importance of
choosing an appropriate super basis. The originally unknown $9\times 9=81$
superexchange coupling constants now become only 6.

\subsection{Superexchange Coupling}

Due to the 3--\textbf{k} symmetry, one can  perform the pair--flip technique
on an arbitrary pair of uranium atoms in the four sublattices and all other
exchange constants can be obtained by permuting their corresponding $x$, $y$ 
$z$ coordinates. There are four equivalent bonds for a pair of uranium sites 
$(i,j)$, so to eliminate double counting one should also divide the obtained
exchange energies by 4 as well as account for any geometric or trigonometric
factors due to the non-collinear order. Since $|\Gamma _{5}\rangle $ ground
state is defined in the pseudospin $J=1$ space, we shall introduce the
reduced density matrix (RDM) as a useful single--particle approximation to
make it compatible with the single--particle based electronic structure
calculation. However the self--consistent ground state of the $UO_{2}$ may
be close to but not equal to the RDMs of the prefect $|\Gamma _{5},-1\rangle 
$ state, so we keep all the calculated results unchanged but replace the
local density matrices of the correlated orbitals by prefect $|\Gamma
_{5},-1\rangle $ RDMs to make our system a well--defined $|\Gamma
_{5}\rangle $ problem. We assume that the multipolar exchange Hamiltonian in
the $J=5/2\oplus 7/2$ single--particle space is built by replacing all
tensor operators, density matrices, and mean values in $J=1$ space to their
corresponding single--particle RDM: $\langle T_{i}^{m}\rangle \rightarrow
\langle \mathscr{T}_{i}^{m}\rangle $, $\langle \rho _{i}\rangle \rightarrow
\langle \mathscr{D}_{i}\rangle $. The single--particle exchange Hamiltonian
shares the same exchange constants as the $J=1$ two--particle version. Two
things to notice here are: 1) the RDM exhibits $J=\frac{5}{2}\oplus \frac{7}{%
2}$ symmetry instead of $J=1$ and this means the rotation from local
coordinates to the global coordinates has to be made in $J=1$ space, else
the pseudospin quasi--particle description will be violated; 2) the RDM
replacement will rescale the length of an operator, $i.e.$ $Tr[\mathscr{T}%
\mathscr{T}^{\dagger }]\neq Tr[TT^{\dagger }]$. Therefore, $\langle %
\mathscr{T}_{i}^{m}\rangle =Tr[\mathscr{D}\mathscr{{T}^{\dagger
m}_{i}}]$ is different from $\langle T_{i}^{m}\rangle =Tr[\rho
T_{i}^{\dagger m}]$. So one has to be cautious when using Eq. (9).

\begin{table}[tbp]
\caption{Comparison between our calculated exchange interaction parameters
using the LDA+U method with $U=4.0$ eV and $J=0.7$ eV and the existing
experimental fits. $C_{0}^{d}$, $C_{0}^{q}$, $K_{0}^{q}$ are in units of
meV, others are dimensionless. Because all the works use different models to
simulate the SL part, there is no appropriate values for them (labeled by
*). Ref.\protect\cite{UE-4} obtained SL via a fully dynamic calculation.
Note also that Ref.\protect\cite{UE-1} assumes the quadrupolar coupling only
comes from SL with real space exchange constant of the 3--$k$ symmetric
form: $K_{ij}^{\Gamma }=K_{0}e^{i\mathbf{q}_{\Gamma }(\mathbf{R}_{i}-\mathbf{%
R}_{j})}$. Ref.\protect\cite{UE-2} only calculates SE part. Their parameters
were obtained via the integrals of Coulomb interaction directly and have no
simple anisotropy form.}%
\begin{ruledtabular}
\begin{tabular}{ccccccc}
Ref. &  $C_{0}^{d}$  & $\chi^{d}_{c}$ & $C^{q}_{0}$ & $\chi^{q}_{c}$ &
$K^{q}_{0}$ & $\chi^{q}_{K}$
\\ \hline
 our work & 1.70 & 0.3 & -3.10 & 0.90 & 2.6 & 1.18 \\
 \cite{UE-4} & 3.1 & 0.25 & 1.9 & 0.25 & $\ast$ & $\ast$  \\
 \cite{UE-1} & 1.25 & 0.8 & 0 & 0 & 0.33 & $\ast$ \\
 \cite{UE-2} & $\sim 1$ & $\ast$ & $\sim 0.1$ & $\ast$ & $\times$ & $\times$
\end{tabular}
\end{ruledtabular}
\end{table}
In Fig. 9, we have plotted the total energies obtained from our LDA+U
calculation. The blue bars are the sum of flipping the multipolar moment at
site $i$ and $j$ individually and the red bars are obtained by flipping both
of them simultaneously. The exchange energy $E^{EX}=\delta
^{2}E_{K_{i},K_{j}}^{Q_{i},Q_{j}}=\delta
E_{K_{i},K_{j}}^{Q_{i},Q_{j}}-\delta E_{K_{i}}^{Q_{i}}-\delta
E_{K_{j}}^{Q_{j}}$ is just the difference between the two bars. One may
notice that the exchange energy of the quadrupoles is much smaller than the
one of the dipoles. This is because the multipolar moments $\langle %
\mathscr{T}_{i}^{m}\rangle $ are about an order smaller than the dipoles.
Once we include this factor, the exchange constants obtained using Eq.(9)
are not necessarily small.

The coupling constants can be simplified by symmetry to the form: 
\begin{equation*}
C_{ij}^{mn}=C^{mn}(\mathbf{R})=C_{0}^{d/q}[1-2(1-\chi _{c}^{d/q})\tau
_{m}\tau _{n}]\delta _{mn},
\end{equation*}%
where $d/q$ means dipole or quadrupole and $\tau =\mathbf{R}/R$ is the
direction vector between $(i,j)$. These constants are shown in Table I,
where the isotropic and anisotropic parts are described by $C_{0}^{d/q}$ and 
$\chi _{c}^{d/q}$ respectively \cite{UE-1}. With the comparison to other
studies, the dipolar part is similar, but the quadrupolar part gives the
opposite result to the past calculations obtained by best fit to experiment 
\cite{UE-3,UE-4}. Not only the anisotropy effect is much smaller, but the
sign is also different which means the quadrupoles tend to be ferromagnetic.
It also means that the SL effects must be as important as SE and their
combination makes the whole system antiferromagnetic. 
\begin{figure}[tbp]
\centering
\includegraphics[width=1.0\columnwidth]{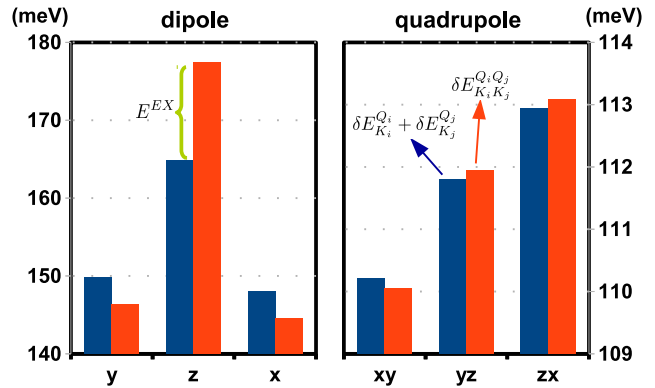}
\caption{(color) Calculted exchange energies using the  LDA+U method with
U=4.0 eV and J=0.7 eV. The ground state energy (when nothing is flipped) is
set to zero. Blue bars correspond to the sum over the energies of flipping a
multipolar moment on site $i$ and $j$ individually. Red bars correspond to
the energy of flipping the multipolar moments on site $i$ and $j$
simultaneously. Exchange energies are the differences between the red and
blue bars (can be positive or negative).}
\end{figure}

\subsection{Spin--Lattice Coupling}

To explain the behavior of the quadrupolar part, we need to include the
effect of dynamic contribution from SL. The coupling between spins and
optical phonons can be written as: 
\begin{equation*}
H_{SL}=\sum_{\mathbf{q} n j}V^{n}(\mathbf{q},j)T^{n}(\mathbf{q})u(\mathbf{q}%
,j),
\end{equation*}%
where $T^{n}(\mathbf{q})=\sum_{\mathbf{R}}T^{n}(\mathbf{R})e^{i\mathbf{q}%
\cdot \mathbf{R}}$, $u(\mathbf{q},j)=[a^{\dagger }(-\mathbf{q},j)+a(\mathbf{q%
},j)]$ and $a^{\dagger }(\mathbf{q},j)$ is the creation operator of a phonon
with wavevector $\mathbf{q}$ in mode $j$. Using the virtual phonon
description, the SL exchange constant of $h^{SL}$ can be approximated as: 
\begin{equation*}
K^{n n}(\mathbf{q})\simeq \sum_{j}\frac{|V^{n}(\mathbf{q},j)|^{2}}{h\omega (%
\mathbf{q},j)}-\varepsilon _{0},
\end{equation*}%
where $\omega (\mathbf{q},j)$ is the phonon frequency and $\varepsilon _{0}$
is the on--site exchange energy which should be subtracted \cite{UE-1}. The
variables $u(\mathbf{q},j)$ and $\omega (\mathbf{q},j)$ have been calculated
in one of our earlier works \cite{ABP-1} and can be fitted to the entire
Brillouin Zone using a simple rigid--ion model \cite{RI-1,RI-2}. If we
further assume\ that the quadrupoles only couple to $t_{2g}^{a}$ and $%
t_{2g}^{b}$ quadrupolar distortions of the O--cage around each U--ion, the
coupling constants are assumed to have the form: 
\begin{equation*}
V^{n}(\mathbf{q},j)=\gamma _{a}\psi _{a}^{n}(\mathbf{q},j)+\gamma _{b}\psi
_{b}^{n}(\mathbf{q},j),
\end{equation*}%
where $\gamma _{a/b}$ are the parameters to be determined, $\psi _{a/b}^{n}(%
\mathbf{q},j)$ are the inner product (projection) between the phonon
distortion $u(\mathbf{q},j)$ and $t_{2g}^{a/b}$ distortion, and $u(\mathbf{q}%
,j)$ can be regarded as the distortion due to a phonon mode \cite{T2G-1}. We
estimate the parameters $\gamma _{a/b}$ by using a coupled frozen--phonon
and frozen--magnon technique: 1) Make a $t_{2g}^{a/b}$ distortion of the
O--cage around an U--ion; 2) Flip a particular tensor component of the
single--ion RDM on a particular site; 3) Calculate the correlation energies: 
$\delta ^{2}E_{a/b}^{mn}=[\delta E_{a/b}^{mn}-\delta E_{a/b}^{0n}-\delta
E^{m0}]$, where the first superscript is the symmetry index of the
quadrupole and the latter index is of $t_{2g}^{a/b}$. So $\delta
^{2}E_{a/b}^{mn}$ is the extra energy of making \textquotedblleft
flip+frozen phonon distortion" simultaneously compared to the energies of
individual \textquotedblleft flip" plus individual \textquotedblleft frozen
phonon distortion"; 4) Then the parameters are roughly: $\gamma _{a}\sim
\delta ^{2}E_{a}^{mn}/\sqrt{2}\langle T^{m}\rangle \psi _{a}^{n}$ and $%
\gamma _{b}\sim \delta ^{2}E_{b}^{mn}/\langle T^{m}\rangle \psi _{b}^{n}$ .
There is a factor $\sqrt{2}$ in $\gamma _{a}$ because when we make the same
displacement of each coordinate component, the length of the total
displacement is $\sqrt{2}$ larger than $t_{2g}^{b}$. By assuming the unit of
phonon vibration about $0.014\mathring{A}$ (as is the static Jahn--Teller
distortion \cite{UE-2}) and making a $t_{2g}$ distortion to be $3\%$ of the
lattice constant, we have: $\gamma _{a}=34meV$ and $\gamma _{b}=48meV$. We
can access nearest--neighbor constants by calculating $K^{n,n}(\mathbf{q},j)$
at $\mathbf{q}=[0,0,0]$ and $\mathbf{q}=\frac{2\pi }{a}[1,0,0]$, and by a
subsequent fit to a cosine function with the on--site exchange energy
assumed to be the average of the curve \cite{UE-1}. We then have: $%
K_{i j}^{m n}=K^{m n}(\mathbf{R})=K_{0}^{q}[1-2(1-\chi _{k}^{q})\tau
_{m}\tau _{n}]\delta _{m n}$ with $K_{0}^{q}=2.6$ meV and $\chi _{k}^{q}=1.18
$.

\begin{figure}[tbp]
\centering
\includegraphics[width=1.0\columnwidth]{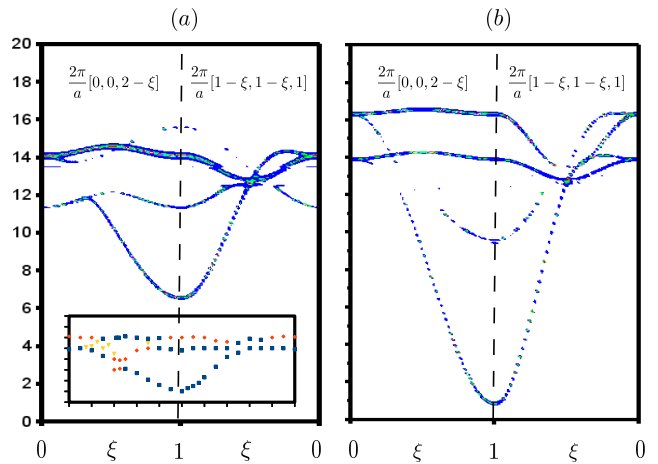}
\caption{(color) Magnetic excitation spectrum for $UO_{2}$ along two
symmetry directions calculated by scanning the colormap of the real part
two--ion susceptibility of our model Hamiltonian \protect\cite{UE-1} with
(a) parameters shown in Table I. (b) The same calculation made by requiring
the overall quadrupole coupling to have 3--$\mathbf{k}$ symmetry: $%
K_{ij}^{\Gamma }=K_{0}e^{i\mathbf{q}_{\Gamma }(\mathbf{R}_{i}-\mathbf{R}%
_{j})}$ with $K_{0}=0.5$ meV \protect\cite{UE-1} in which case the
anisotropy gap is greatly reduced. Bottom inset: data from inelastic neutron
scattering experiments plotted in the same $x-y$ scale\protect\cite{SW-1}.
Triangles (yellow) are measured in a direction set by a reciprocal lattice
vector\protect\cite{SW-2}. Rhombus (orange) are weaker cross sections.}
\end{figure}

\subsection{Magnetic Excitation Spectrum}

Combined with the superexchange contribution and using the Green function
method with random phase approximation \cite{UE-1}, we calculate the
magnetic excitation spectrum of UO$_{2}$ that is shown in Fig. 10. We find
that the values and the characteristics of our results are basically in
agreement with experiment. The major difference is the disappearance of
anti--crossing at a few $\mathbf{q-}$points and much larger anisotropy (gap)
at the $X$--point. The disappearance of the anti--crossing is reasonable
because it comes from the coupling between magnon and phonon branches. As
for the overestimated anisotropy at the $X$--point, it is believed to come
from the oversimplified SL model in our calculation. We have plotted the
spin/quadrupolar wave spectrum by enforcing the overall quadrupole coupling
to have 3--$k$ symmetry as Ref. \cite{UE-4} with the parameter $K_{0}=0.5$
meV (which is almost the same value as our isotropic part) and it gives a
much smaller gap which fits the experiment well (see Fig. 10). It
demonstrates that a SL model which makes the whole quadrupolar coupling to
have 3--$k$ symmetry will be helpful in fitting the experiment but, in this
case, the simplified form of our model will be also lost.

\section{Conclusion}

In conclusion, we have introduced the framework of multipolar operators and
the benefits of using them as a language to describe the exchange
interactions in spin--orbit coupled systems. We have also developed a method
to calculate the exchange constants via a density functional based total
energy calculation. With its application to $UO_{2}$, the superexchange
tends to have ferromagnetic quadrupolar coupling rather than
antiferromagnetic one which is very different from the past reports using
best fits to experiments. It demonstrates that our method has the potential
to explore magnetic spin--orbit coupled systems in more details. As for the
spin--lattice interaction, we have performed a very similar calculation to
estimate their couplings and the overall behavior is accounted for by the
competition between the superexchange and spin--lattice counterparts. An
accurate description of spin--lattice interactions and applications to
hidden order systems would be beneficial for the future work.

\section{Acknowledgement}

We are grateful to X. Wan and R. Dong for their helpful discussions. This
work was supported by U.S. DOE Nuclear Energy University Program under
Contract No. 00088708.


\end{document}